# 24-1: Detection of Ionizing Radiation by Plasma-Panel Sensors: Cosmic Muons, Ion Beams, and Cancer Therapy

## Peter S. Friedman
### Integrated Sensors, LLC, Ottawa Hills, Ohio, USA

## Robert Ball, J. W. Chapman, Claudio Ferretti, Daniel S. Levin, Curtis Weaverdyck, Bing Zhou
### University of Michigan, Department of Physics, Ann Arbor, Michigan, USA

## Yan Benhammou, Erez Etzion, Nir Guttman, M. Ben Moshe, Yiftah Silver
### Tel Aviv University, Raymond and Beverly Sackler School of Physics and Astronomy, Tel Aviv, ISRAEL

## James R. Beene and Robert L. Varner Jr.
### Oak Ridge National Laboratory, Holifield Radioactive Ion Beam Facility, Oak Ridge, Tennessee, USA

## Abstract
*The plasma panel sensor is an ionizing photon and particle radiation detector derived from PDP technology with high gain and nanosecond response. Experimental results in detecting cosmic ray muons and beta particles from radioactive sources are described along with applications including high energy and nuclear physics, homeland security and cancer therapeutics.*

## 1. Introduction

The plasma panel sensor (PPS) is a new radiation detector technology being developed for a number of scientific and commercial applications [1]-[6]. The PPS (see Fig. 1), which is based on the plasma display panel (PDP), is designed to leverage off of the low cost consumer electronics PDP technology developed for HDTV. PDPs comprise millions of cells per square meter, each of which when provided with a signal pulse can initiate and sustain a plasma discharge. However, rather than the plasma discharge being initiated *externally* by a signal from a driver chip (i.e. address pulse) as in a PDP, the PPS discharge is initiated *internally* by an ionization event created within the device by an ionizing photon or particle interacting with the detector. In other words the order of processes is reversed, instead of applying voltage to produce light emission via a plasma discharge, we detect the plasma discharge generated by ionizing radiation entering a PPS cell.

The PPS was initially conceived to be able to leverage off of the mature PDP technology base with its low cost manufacturing infrastructure, by using similar materials and manufacturing processes [1]. Thus in addition to offering the possibility of using inexpensive materials and fabrication processes for the production of highly pixelated, high performance devices (e.g. PDPs cost ~ \$0.20 per sq. inch), the PPS offers a number of other potential advantages including: pulse rise times of ~ 1-2 ns and FWHM response times less than 5 ns (see Fig. 2), high gain (e.g. higher than photomultiplier tubes), a thin and compact/portable flat-panel structure with very low mass and a hermetic seal eliminating the need for a gas flow system, and a materials composition that is inherently radiation damage resistant (e.g. glass substrates, metal electrodes and stable gas mixtures). These potential attributes of the PPS are attracting significant interest for applications ranging from: detection of nuclear materials (e.g. U and Pu) for homeland security [2], detecting minimum ionizing particles (MIPs) at the Large Hadron Collider (i.e. LHC at CERN) [3]-[6], radioactive ion beams monitors for nuclear physics (e.g. DOE and Oak Ridge National Laboratory), proton beam detectors for improved radiation therapeutics in treating cancer, medical imaging, etc.

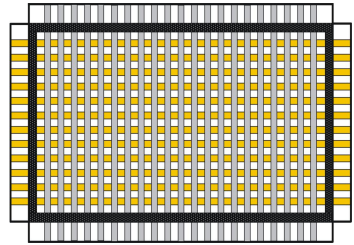

Figure 1 (Top) – Drawing of 2-electrode, columnar discharge PPS structure. (Bottom) – Photograph of 2-electrode, modified-PDP columnar-discharge panel used for experiments in Figures 2-6.

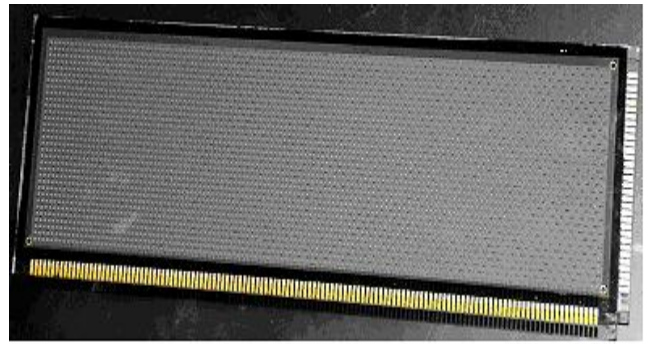

## 2. Discussion and Results

The PPS can be thought of as a dense array of micro-Geiger cells having a discharge gap on the order of 100-400 μm with the drift field (i.e. gas gap) on the order of 0.5 to 1mm; however, these are ballpark values and can easily vary by a factor of two or more, as they are application dependent. For example, betas and protons cab be highly ionizing, depending upon their energy; whereas muons are *minimum* ionizing particles (MIPs) and so require a much larger gas path to generate an equivalent number of gas discharge events.

The active area in the panel is the gas volume between the electrodes, which is enclosed by a glass substrate. In order to determine the response to radiation we used GEANT4 to simulate the energy loss and scattering occurring in the glass substrate prior to entering the panel gas discharge region. We have simulated the energy spectrum of betas entering the active pixel region, based on the known energy originally emitted by both the $^{90}Sr$ and $^{106}Ru$ sources. Most of our efforts have focused on the response of modified-PDPs/PPS devices, that produce signals when exposed to radioactive sources or when being traversed by a cosmic muon.

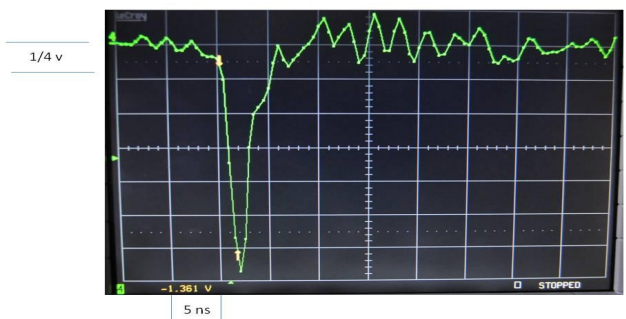

**Figure 2 – Gas discharge pulse from 2-electrode PPS with 1% $CO_2$ in 99% Ar, at 600 torr and operating at 840V. The experiment employed a $^{106}Ru$ beta-source in conjunction with a triple coincidence hodoscope arrangement (i.e. trigger). Rise time was ~ 1 ns (20%-80%) and < 2 ns for 10%-90%, with pulse duration (FWHM) of 1.9 ns.**

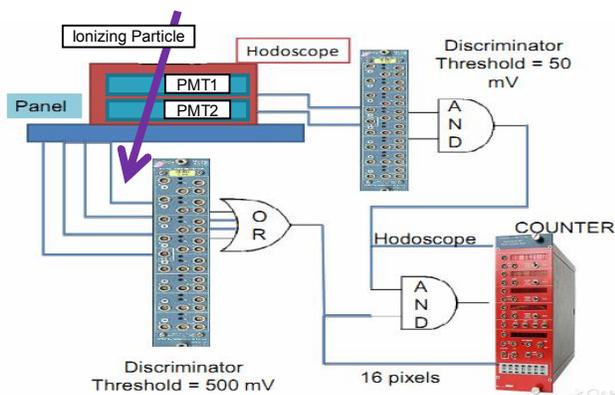

**Figure 3 – Hodoscope measurement setup for both cosmic ray muons and $^{106}Ru$ beta particles.**

In order to explore the behavior of the PPS devices under various kinds of radiation, we have constructed two test benches: one at the University of Michigan and the other at Tel Aviv University. Each test bench includes a gas delivery system, a triggering system, and a data acquisition (DAQ) system. The triggering is being done with a hodoscope (see Fig. 3) that includes a set of scintillation pads. The DAQ is for characterization of the signal induced in the panel during discharge. To accomplish this we are using two sets of 5 GHz digitizer boards (i.e. digital sampling oscilloscope) based on the DRS4 chip developed at the Paul Scherrer Institut (http://drs.web.psi.ch/). For the discharge rate measurements we are using a set of discriminators and counters (see Fig. 3). With the two digitizers (four channels each) we are able to read a 4 x 4 array of pixels simultaneously, thus achieving a 2D position measurement of radiation traversing the panel. We are transitioning to an array of 24 x 24 pixels in our new DAQ.

We are investigating the panel response to radiation with various gases at different pressures. The gas pressures range from ~ 200 to 700 torr. The signals we observe from all of the gases tested have large amplitudes of at least several volts, so there is no need for amplification electronics. For each gas the shape of the induced signals are uniform. The leading edge rise time is typically a few nanoseconds (see Fig. 2). The discharge spreading to neighboring pixels is gas dependent, but has been measured, for example, to be ~ 2% for an Ar/$CO_2$ mixture. A typical example is shown in Fig. 4 which is a single pixel response to a $^{106}Ru$ beta source, with no response seen on the neighboring electrodes (i.e. the adjacent electrodes/channels are shown in different colors).

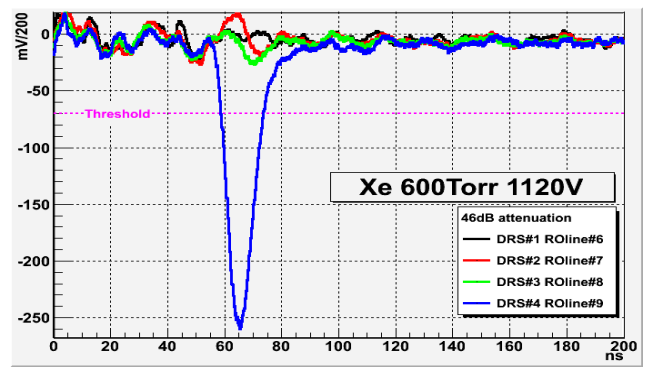

**Figure 4 – Gas discharge pulse (i.e. "blue" readout line #9; 46db attenuation) from PPS with 100% Xe at 600 torr. Beta source was $^{106}Ru$, used in conjunction with a triple hodoscope trigger arrangement. The adjacent anodes (i.e. channels 6, 7 & 8) appear as the black, red and green lines, and show no indication of discharge spreading.**

Cosmic ray muons allow us to test the panel's response to minimally ionizing particles (MIPs). Using the setup described above we are able to associate signals induced in the panel with cosmic muons. With $CF_4$ gas at 600 torr we have measured the panel total efficiency to be ~ 10% for a voltage range of more than 50 volts. The total efficiency is defined as the ratio of signals in the panel that coincide with the trigger versus the total number of triggers (i.e. all of the triggers from the hodoscope are associated with cosmic ray muons). When taking into account that only the pixel area itself is active, it yields that per pixel the efficiency to detect muons (with $CF_4$ gas at 600 torr and a 450 μm gap) should be much higher at ~ 80% to 90%. For this panel with a scintillator trigger we have also measured the multi-pixel response to cosmic ray muons and see evidence of a broad efficiency plateau of ~ 80 volts (see Fig. 5). We have also measured the time of the signal crossing the threshold with respect to the scintillator trigger; the corresponding distribution for 197 cosmic muons (in a panel filled with $SF_6$ at 200 torr and 1530 volts) is nicely fit by a Gaussian with a width (σ) of less than a 5 ns section.

We have tested these panels with several beta sources, including $^{90}Sr$, $^{106}Ru$ and $^{137}Cs$, as well as with cosmic muons and observe a high probability for detecting "hits" from such ionizing particles when they enter the voxel space defined by the cell discharge gap volume dimensions. To take advantage of this, new cell structures are being designed to maximize the effective cell active discharge region to maximize the device efficiency. We are also changing our device fabrication process and the cell design to significantly improve pixel uniformity and thereby maximize the operational range for the panel. We do observe that relatively little discharge spreading appears be occurring from a discharging pixel to its neighbors (see Fig 4) and therefore does not appear to be a problem in the "open structure" panels under investigation.

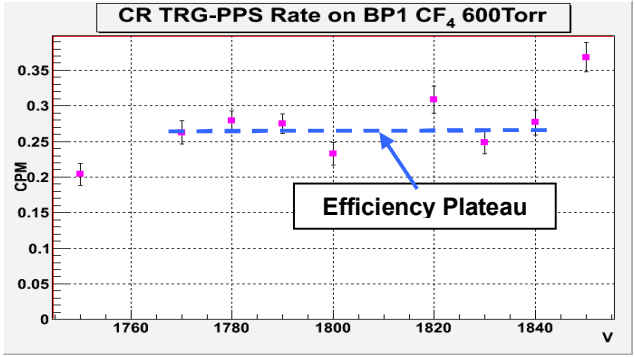

**Figure 5 – Count rate (cpm) vs. voltage for triggered cosmic muon "hits" over a 5 cm² area in PDP filled with CF₄ at 600 torr. Efficiency plateau is about 80 volts.**

We have operated the PPS over a wide range of gas pressures using a variety of discharge gases including: $Ar+CO_2$, $Ar+CF_4$, $CF_4$, $SF_6$ and Xe. Not unexpectedly, the device performance has been shown to be very much gas dependent, with the breakdown voltages varying by more than 1000 volts for different gas mixtures in the same panel. The discharge spreading to neighboring cells is also very much gas dependent, yet we have shown that gas discharges can be confined to a single cell, with several gas mixtures showing minimal, if any, gas discharge spreading to adjacent cells. We consider it very significant that in an "open" cell structure, we have demonstrated minimal discharge spreading, especially given that our devices operate in the Geiger mode, producing large amplitude, high gain discharges. The fact that this has been done *without* an internal barrier structure around each cell is particularly encouraging. But equally important is that unlike most other gaseous detectors, we have not had to add a hydrocarbon quenching gas component that would certainly degrade in a plasma discharge environment. The elimination of hydrocarbon quenching gases is considered critical to realizing a stable, hermetically-sealed PPS device, without the cost, bulk and complication of having to constantly flush the gas as required in most other position sensitive gaseous detectors.

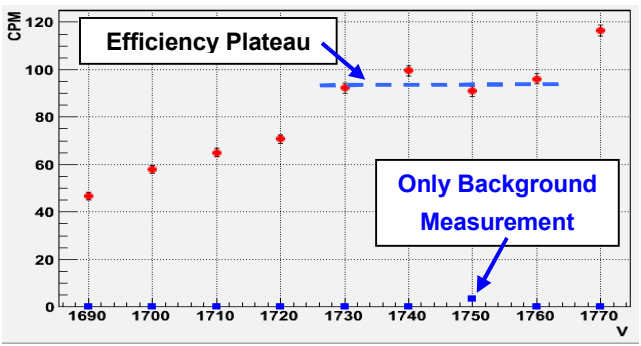

**Figure 6 – Hit rate detected by single pixel from ⁹⁰Sr source. Background rate in "blue" was "0" everywhere except 1750V.**

The panels tested appear capable of stable operation in a sealed PPS gas environment (i.e. without gas flow) with low background counts/noise (e.g. background signals of less than 0.5%, see Fig

6). We have also been able to demonstrate 2D position capability by using both the sense and HV electrodes, with potentially high X-Y resolution for small cell dimensions. In fact for a panel with a 2.5 mm pixel pitch, we can resolve the position of a ¹⁰⁶Ru beta source (behind a 1.2 mm slit) to within ~ 1 mm. For a given panel structure and gas, the discharge signals look remarkably uniform and are inherently digital. For the devices tested, we estimate the PPS gain to be at least $10^7$. We expect faster discharge times in the sub-nanosecond range as we transition to smaller cell sizes, better cell physical and electrical isolation, and lower panel capacitance. We believe that the fast rise times and short pulse durations are largely due to the very high gain of the PPS Geiger-mode electron avalanche, which might be generated via a limited *micro*-streamer mode mechanism (e.g. Limited Streamer Tubes).

We have developed a practical modeling and simulation capability to: (1) provide better theoretical insight into the device physics and a clearer understanding of the interplay between the various device design parameters, materials selection and electronics readout design; and (2) to provide important design guidance for device optimization with respect to specific applications and to better understand the various performance tradeoffs associated with each particular device design. Our approach starts with a simplified schematic of a single PPS discharge cell. We then created a more realistic model and schematic of the discharge cell that includes stray capacitances, line resistance, and self-inductance. The parameters were determined from a COMSOL electrostatic model. Finally we expanded the single cell to a chain of cells by adding in the neighboring cells to form a larger array system. Represented in this expanded cell array/schematic are the embedded cell resistances, the cell capacitances, stray capacitances, self-inductances and the termination resistance. The various capacitive couplings were modeled with COMSOL. COMSOL-3D was employed to model the electric field and the charge motion inside the pixels, and the electronic properties of the different components (e.g. capacitances and inductances of the cells). SPICE was employed to simulate the electrical characteristics of the signal induced in the panel during discharge. The parameters in the SPICE models were determined with our COMSOL electrostatic model. The full SPICE model connects all of the neighboring cells into a single matrix to form a large cell array or a small panel sector. For example, we are now able to superimpose measured (i.e. experimental) signals over the SPICE simulations, with the result being an excellent match of the basic discharge shape. By testing the influence of the various parameters, we are able to enhance our understanding of how these devices operate, their performance advantages and limitations, and how they can be optimized for specific applications.

Recently we demonstrated the ability to detect 226 MeV protons from a medical accelerator used in proton therapy for the treatment of cancer. We were able to demonstrate position sensitivity, which could prove important for proton imaging of tumors in real-time, and also the potential to do proton dosimetry in a future PPS designed for reduced capacitance. In our first proton beam test we were able to follow and accurately resolve the location to within one sense electrode for both a 1 mm and a 10 mm diameter beam as we translated the beam relative to the PPS a distance of several centimeters, and we were able to do this under a proton flux of more than 2 x 10⁶ protons per second distributed over an area of a few square cm.

For homeland security applications, we have investigated the possibility of fabricating thermal neutron detectors (e.g. as a replacement for [3]He detectors) based on incorporating a thin-film layer(s) or thin-foil of gadolinium *within* the PPS. The key to replacing [3]He with Gd is to construct a "gamma-blind" electron detector that can detect the Gd conversion electrons (note Gd has a thermal neutron capture cross-section that is unparalleled among stable elements). We feel than an ultra-low-mass Gd-foil based PPS might be such a detector in providing a potentially highly efficient, nearly gamma-blind detector of the conversion electrons. We simulated such a device using GEANT4, which resulted in a maximum efficiency estimate and a gamma-neutron discrimination ratio that was close to the values set by the U.S. Domestic Nuclear Detection Office for [3]He replacement neutron detectors [2]. An additional advantage is that the device would make an extremely light weight, compact package suitable for portable applications.

In summary the open-cell PPS structure has been shown to be able to confine the discharge to a single cell, while achieving response times on the order of a few nanoseconds or faster. For higher resolution panels with smaller cell dimensions, we anticipate device response times in the *sub-nanosecond* range. Key objectives of our initial experimental program were to demonstrate that: 1) PPS devices can be fabricated as high gain, micropattern detectors and successfully operated beyond the proportional region and above the gas breakdown voltage (i.e. as a Geiger-mode type device) with high performance capability; 2) discharges self-terminate and can be self-contained to yield high spatial and high temporal resolution; 3) low cost, commercial PDP technology can be modified to detect ionizing radiation; 4) signals have fast discharge times and large amplitudes; 5) hermetically-sealed PPS gas devices appear to be stable; 6) useful models can be constructed with simulations that can be experimentally verified to confirm and enhance our understanding of how these devices operate, their performance advantages and limitations, and how they can be optimized for specific applications.

We are gratified that all six (6) of the initial program objectives have been confirmed and we are now moving to focus on specific device applications and commercialization. We believe that we have been able to demonstrate the viability, merit and potential capability of the PPS as a hermetically-sealed, high gain, rad-hard detector with both high spatial and high temporal resolution, high rate capability and low cost.

## 3. Impact

The potential impact of the PPS radiation detector technology includes a broad range of commercial applications. In this paper we have focused primarily on ionizing particle detection, but we are also pursuing medical applications for ionizing photon detection, such as X-ray radiation therapeutics. Our collaborators at Oak Ridge National Laboratory have run simulations using GEANT4 on the PPS configured for X-ray detection and have found that our devices should be able to measure the incident beam in real time as the patient is being treated with very little scattering of the beam to the patient. In terms of radiation therapeutics, we have now shown that the PPS is capable of detecting proton beams in the energy range used for treating cancer. The detection of MIPs has been discussed in some detail and is of critical importance to high energy physics. Similarly the detection of radioactive ion beams is of importance to nuclear physics. The PPS should also be capable of detecting neutrons [2] emitted by fissile materials such as U and Pu, which has important implications for homeland security. In the future we plan to also explore applications for medical imaging such as PET, CT, SPECT, etc. Given the breath of possible applications and the cost advantages of the PPS technology, the commercial impact and potential benefits of this technology could have a large impact on a number of important fields. Finally radiation detectors for homeland security and various medical applications constitute a multibillion dollar business opportunity. And unlike flat panel displays which are consumer electronics, *the PPS can sell for one to two orders of magnitude above its manufacturing price and still be priced below competing radiation detector technologies.*

## 4. Acknowledgements

This work is being supported in part by the U. S. Department of Energy under Grant Numbers: DE-FG02-07ER84749, DE-SC0006204, and DE-SC0006219. This work is also supported in part by the Office of Nuclear Physics at the U. S. Department of Energy, and by the United States–Israel Binational Science Foundation under Grant No. 2008123. We would also like to thank Dr. Hassan Bentefour at IBA for making their proton beam accelerator facility available for testing our devices.